\documentclass[aps,prl,twocolumn,floatfix]{revtex4}

\usepackage{amsmath,amssymb,bm}
\usepackage{epsfig}
\usepackage{graphics}

\begin{document}

\bibliographystyle{prsty}

\title
{Variational Minimization of Orbital-dependent Density Functionals}
\author{Cheol-Hwan Park$^{1,2}$}
\email{chpark77@mit.edu}
\author{Andrea Ferretti$^{2,3,4}$}
\author{Ismaila Dabo$^{5}$}
\author{Nicolas Poilvert$^{1}$}
\author{Nicola Marzari$^{1,2}$}
\affiliation{$^1$Department of Materials Science and Engineering, Massachusetts
Institute of Technology, Cambridge, Massachusetts 02139, USA\\
$^2$Department of Materials, University of Oxford, Oxford OX1 3PH, UK\\
$^3$INFM-S3 \& Physics Department, University of Modena and Reggio Emilia,
Modena, Italy\\
$^4$Centro S3, CNR--Instituto Nanoscienze, I-41125 Modena, Italy\\
$^5$Universit\'e Paris-Est, CERMICS, Project Micmac ENPC-INRIA,
6 \& 8 avenue Blaise Pascal, 77455 Marne-la-Vall\'ee Cedex 2, France}
\date{\today}

\begin{abstract}
Density-functional theory has been one of the most successful approaches ever to 
address the electronic-structure problem; nevertheless, since its 
implementations are by necessity approximate, they can suffer from a number of fundamental 
qualitative shortcomings, often rooted in the remnant electronic self-interaction 
present in the approximate energy functionals adopted.
Functionals that strive to correct for such self-interaction errors, such as those
obtained by imposing the Perdew-Zunger self-interaction
correction [Phys. Rev. B {\bf 23}, 5048 (1981)] 
or the generalized Koopmans' condition
[Phys. Rev. B {\bf 82}, 115121 (2010)], 
become orbital dependent or orbital-density dependent, and provide a very promising avenue to 
go beyond density-functional theory,
especially when studying electronic, optical and
dielectric properties, charge-transfer excitations, and molecular dissociations.
Unlike conventional density functionals, these functionals 
are not invariant under unitary transformations of occupied electronic
states, which leave the total charge density intact, and this added complexity
has greatly inhibited both their development and their practical applicability.
Here, we first recast the minimization
problem for non-unitary invariant energy functionals
into the language of ensemble density-functional theory [Phys. Rev. Lett. {\bf 79}, 1337 (1997)],
decoupling the variational search into an inner loop of unitary
transformations that minimize the energy at fixed orbital subspace, and an outer-loop evolution of 
the orbitals in the space orthogonal to the occupied manifold.
Then, we show that the potential energy surface in the inner loop is far from convex parabolic
in the early stages of the minimization
and hence minimization schemes based on these assumptions are unstable,
and present an approach to overcome such difficulty. 
The overall formulation allows for a stable, robust, and
efficient variational minimization of non-unitary-invariant functionals, essential to
study complex materials and molecules, and to investigate the bulk thermodynamic limit,
where orbitals converge typically to localized Wannier functions. 
In particular, using maximally localized Wannier functions as an initial guess
can greatly reduce the computational costs needed to reach the energy minimum
while not affecting or improving the convergence efficiency.
\end{abstract}
\maketitle

\section{I. Introduction}

Density functional theory (DFT)~\cite{HohenbergKohn64,KohnSham65}
has become the basis of much computational materials
science today, thanks to its predictive accuracy in describing 
ground-state properties directly from first principles.
While DFT is in principle exact, in any practical implementation
it requires an educated guess for the exact form of the energy functional.
For many years, local or semi-local approximations to the exchange-correlation
energy, such as the local density approximation
(LDA)~\cite{CeperleyAlder80,perdew_zunger}
or the generalized gradient approximation~\cite{PerdewBurkeErnzerhof96}
have been
successfully applied to a wealth of different systems~\cite{marzari_mrs}.
Still, these
approximations lead to some dramatic failures, including
the overestimation of dielectric response, incorrect
chemical barriers for reactions involving strongly-localized orbitals~\cite{kulik:prl,zhou:prb},
energies of dissociating molecular species, and
excitation energies of charge-transfer complexes,
to name a few~\cite{cohen_insights_2008}.

Key to these failures is the
self-interaction error of approximate
DFT~\cite{cohen_insights_2008,perdew_zunger},
where the electrostatic and exchange-correlation contributions
to the effective energy of the entire charge distribution are
not ``purified'' from this spurious self interaction of an
individual electron with itself.
To address this issue,
Perdew and Zunger (PZ) introduced first an elegant solution to this problem,
where a self-interaction correction (SIC) is added to the total energy calculated from 
approximate DFT (e.g. within the LDA~\cite{CeperleyAlder80,perdew_zunger}),
but practical applications have remained
scarce~\cite{svane:prl,hughes_lanthanide_2007,stengel_spaldin,capelle,kuemmel,Baumeier,Filippetti,ruzsinszky_density_2007,suraud,sanvito}.

An important property of DFT with local or semi-local exchange-correlation functionals
is the invariance of the total energy with respect to unitary transformation
of the occupied electronic states.  However, SIC-DFT does not 
have this invariance property, and in fact finding the
optimal unitary transformation
given a set of orbital wavefunctions is crucial to the numerically
  consistent minimization of density functionals with
SIC~\cite{pederson_local-density_1984,svane:1996,svane:2000,goedecker_umrigar,vydrov_scuseria,stengel_spaldin,klupfel}.
In this paper, we focus on the variational minimization of
energy functionals that do not satisfy unitary invariance
in order to provide a stable, robust, and efficient 
determination of the electronic structure in this 
challenging case.  In particular,
we adopt the formulation of ensemble DFT~\cite{eDFT} to decouple the 
variational minimization into an inner loop of unitary
transformations and an outer loop of evolution for the occupied manifold, and
suggest optimal strategies for the dynamics of unitary transformations.
In the solid-state limit, this dynamics gives rise to a localized Wannier representation
for the electronic states, and we assess their relation with 
maximally localized Wannier functions
(MLWFs)~\cite{marzari_vanderbilt,souza_marzari_vanderbilt}
as obtained in the absence of SIC.

The remainder of the paper is organized as follows.
In Sec.~II, DFT with SIC is briefly reviewed, the method
of inner-loop  minimization is explained,
and the issue of using MLWFs as an initial guess
for the wavefunctions is discussed.
In Sec.~III, we present and discuss the results.
First, we present results on how the total energy varies with
the unitary transformation of the occupied electronic states.
Second, we discuss the stability and efficiency of our method
for inner-loop minimization.
Finally, we show how the calculated total energy converges
both as a function of the outer-loop iterations and
as a function of the CPU time and discuss the optimal
scheme for total energy minimization of energy functionals with SIC.
We then summarize our findings in Sec.~IV.

\section{II. Methodology}
\subsection{A. Background}

For simplicity, we consider in the following the wavefunctions to be real;
however, the discussion can straightforwardly be extended to
complex wavefunctions.
The total energy of the interacting electron system from Kohn-Sham DFT
within the LDA is given by~\cite{KohnSham65}
\begin{eqnarray}
&&E_{\rm LDA}[\{\psi_{\sigma i}\}]\nonumber\\
&&=-\sum_{\sigma}\sum_{i=1}^N\frac{1}{2}\int\psi_{\sigma i}({\bf r})\nabla^2
\psi_{\sigma i}({\bf r})\,d{\bf r}
+\int V_{\rm ext}({\bf r})\rho({\bf r})d{\bf r}
\nonumber\\
&&+\frac{1}{2}\int\int\frac{\rho({\bf r})\rho({\bf r}')}{|{\bf r}-{\bf r}'|}\,
d{\bf r}\,d{\bf r}'
+\int\epsilon_{\rm xc}^{\rm LDA}(\rho({\bf r}))\,\rho({\bf r})d{\bf r}\,,
\label{eq:E_LDA}
\end{eqnarray}
where $\sigma$ is the spin index, the band index $i$ runs through
the $N$ occupied electronic states, and
$\rho({\bf r})=\sum_{\sigma}\sum_{i=1}^N|\psi_{\sigma i}({\bf r})|^2$
is the total charge density.
The first term on the right hand side of Eq.~(\ref{eq:E_LDA}) is
the kinetic energy, the second term the interaction energy between electrons
and the ion cores, the third term the Hartree interaction energy,
and the last term the exchange-correlation energy.
This energy functional $E_{\rm LDA}[\{\psi_{\sigma i}\}]$
is invariant under the following unitary transformation
\begin{equation}
\psi'_{\sigma i}({\bf r}) = \sum_{j=1}^N\psi_{\sigma j}({\bf r})\, O_{\sigma ji}
\label{eq:U}
\end{equation}
for an arbitrary unitary matrix $O_{\sigma}$
since the total charge density $\rho({\bf r})$
and the kinetic energy [Eq.~(\ref{eq:E_LDA})] are
invariant under this transformation.
Given that the wavefunctions are real, we consider $O_\sigma$ to be an
orthogonal matrix, i.\,e.\,,
real and satisfying $O_\sigma^{\rm t}O_\sigma=I$ where $I$
is the $N\times N$ identity matrix.

For some density functionals with SIC~\cite{perdew_zunger,dabo:NK},
the total energy $E_{\rm total}[\{\psi_{\sigma i}\}]$ is given by
\begin{equation}
E_{\rm total}[\{\psi_{\sigma i}\}]=E_{\rm LDA}[\{\psi_{\sigma i}\}]
+E_{\rm SIC}[\{\rho_{\sigma i}\}]\,,
\label{eq:E_total}
\end{equation}
where $\rho_{\sigma i}({\bf r})=|\psi_{\sigma i}({\bf r})|^2$.
$E_{\rm SIC}[\{\rho_{\sigma i}\}]$ and hence
$E_{\rm total}[\{\psi_{\sigma i}\}]$
are in general not invariant under orthogonal transformations
because they are dependent not only on the total charge
density $\rho({\bf r})$, which is invariant under orthogonal
or unitary transformation, but also on the charge densities,
$\rho_{\sigma i}({\bf r})$'s,
arising from different orbitals.

This can be seen by considering how the SIC energy varies
under the orthogonal transformation of Eq.~(\ref{eq:U}).
To this end, it is useful to recall that
an orthogonal matrix $O_{\sigma}$ can be written as
\begin{equation}
O_{\sigma}=e^{A_\sigma}
\label{eq:O_eA}
\end{equation}
where $A_\sigma$ is an antisymmetric matrix;  if we further consider
the case where
the norm of $A_\sigma$ is much less than that of an identity matrix,
we can assume
\begin{equation}
O_{\sigma}\approx I+A_\sigma\,.
\label{eq:O_eA2}
\end{equation}
Therefore, the transformed wavefunctions are given by
\begin{equation}
\psi'_{\sigma j}({\bf r})\approx\psi_{\sigma j}({\bf r})+\sum_{i=1}^N\psi_{\sigma i}({\bf r}) A_{\sigma ij}\,,
\label{eq:A}
\end{equation}
from which
\begin{equation}
\frac{\partial\rho_{\sigma j}({\bf r})}{\partial A_{\sigma ij}}
=2\psi_{\sigma j}({\bf r})\psi_{\sigma i}({\bf r})\,,
\label{eq:drhodA}
\end{equation}
and (using the antisymmetry of $A_\sigma$)
\begin{equation}
\frac{\partial\rho_{\sigma i}({\bf r})}{\partial A_{\sigma ij}}
=-2\psi_{\sigma j}({\bf r})\psi_{\sigma i}({\bf r})\,.
\label{eq:drhodA2}
\end{equation}
Finally, if we define the SIC potential
\begin{equation}
v^{\rm SIC}_{\sigma i}({\bf r})=\frac{\delta E_{\rm SIC}}{\delta \rho_{\sigma i}({\bf r})}\,,
\label{eq:vsic}
\end{equation}
we obtain the gradient of SIC energy with respect to the
transformation matrix elements
\begin{eqnarray}
&&G_{\sigma ij}
\equiv\frac{\partial E_{\rm SIC}}{\partial A_{\sigma ij}}\nonumber\\
&&=2\int \psi_{\sigma i}({\bf r})
\left[v^{\rm SIC}_{\sigma j}({\bf r})-v^{\rm SIC}_{\sigma  i}({\bf r})\right]
\psi_{\sigma j}({\bf r})\,d{\bf r}\,,
\label{eq:pederson}
\end{eqnarray}
which is a result originally obtained by
Pederson {\it et al.}~\cite{pederson_local-density_1984}.
Note that this gradient matrix $G_\sigma$ is also antisymmetric,
just like $A_\sigma$.
Therefore, at an energy minimum, the wavefunctions satisfy
\begin{equation}
0
=\int \psi_{\sigma i}({\bf r})
\left[v^{\rm SIC}_{\sigma j}({\bf r})-v^{\rm SIC}_{\sigma  i}({\bf r})\right]
\psi_{\sigma j}({\bf r})\,d{\bf r}\,,
\label{eq:pederson2}
\end{equation}
which was referred to as the ``localization condition''
by Pederson {\it et al.}~\cite{pederson_local-density_1984}.

To date, the most widely used SIC scheme is PZ
SIC~\cite{perdew_zunger} (and its few refinements, e.\,g.\,,
Refs.~\cite{Filippetti,lundin_eriksson,davezac}).
In PZ scheme, the SIC energy is given by
\begin{eqnarray}
E^{\rm PZ}_{\rm SIC}[\{\rho_{\sigma i}\}]&=&-\sum_{\sigma}\sum_{i=1}^N
\frac{1}{2}\int\int\frac{\rho_{\sigma i}({\bf r})\rho_{\sigma i}({\bf r}')}{|{\bf r}-{\bf r}'|}\,
d{\bf r}\,d{\bf r}'\nonumber\\
&-&\sum_{\sigma}\sum_{i=1}^N\int\epsilon_{\rm xc}^{\rm LDA}(\rho_{\sigma i}({\bf r}))\,\rho_{\sigma i}({\bf r})d{\bf r}\,.
\label{eq:E_SIC_PZ}
\end{eqnarray}
The rationale underlying PZ SIC is both simple and beautiful:
correcting the total energy by subtracting
the incorrect energy contributions from the interaction of an electron
with itself --- i.\,e.\,, the Hartree, exchange, and
correlation energies. Hence PZ SIC is exact for one-electron
systems, or in the limit where the total charge density can
be decomposed into non-overlapping one-electron charge density contributions.

Recently, an alternative scheme suitable for many-electron systems
based on the generalized Koopmans condition~\cite{koopmans}
was introduced in Ref.~\cite{dabo:NK}.
In brief, one could start from Janak's theorem~\cite{janak}
that states that in DFT
the orbital energy $\epsilon_{\sigma i}(f)$ with fractional occupation
of a state being $f_{\sigma i}=f$ is
\begin{equation}
\epsilon_{\sigma i}(f)=\left.\frac{dE_{\sigma i}(f')}{df'}\right|_{f'=f}\,,
\end{equation}
where $E_{\sigma i}$ is the Kohn-Sham total energy minimized under the
constraint $f_{\sigma i}=f$. If there were no self-interaction,
the orbital energy of a state $\epsilon_{\sigma i}(f)$ would
not change upon varying its own occupation $f$.
In other words, for a self-interaction-free functional,
\begin{equation}
\epsilon_{\sigma i}(f)={\rm constant}\,\,\,(0\le f\le1)\,.
\label{eq:NK_condition}
\end{equation}
Alternatively, using Janak's theorem~\cite{janak},
this can be rewritten as
\begin{eqnarray}
\Delta E^{\rm Koopmans}_{\sigma i}(f)\equiv
E_{\sigma i}(f_{\sigma i})-E_{\sigma i}(0)
=f_{\sigma i}\,\epsilon_{\sigma i}(f)\nonumber\\
(0\le f\le1)\,,
\label{eq:delta_E}
\end{eqnarray}
which is equivalent to the generalized Koopmans theorem~\cite{dabo:NK},
telling us that the total energy
varies linearly with the fractional occupation $f_{\sigma i}$.
In conventional DFT, however, Eq.~(\ref{eq:NK_condition})
or Eq.~(\ref{eq:delta_E}) does not hold and instead,
\begin{equation}
\Delta E_{\sigma i}\equiv
E_{\sigma i}(f_{\sigma i})-E_{\sigma i}(0)
=\int_0^{f_{\sigma i}}\epsilon_{\sigma i}(f')\,df'\,.
\label{eq:delta_E2}
\end{equation}
From Eqs.~(\ref{eq:delta_E}) and~(\ref{eq:delta_E2}),
the non-Koopmans (NK) energy $\Pi_{\sigma i}(f)$ -- i.\,e.\,,
the deviation from the linearity
for the energy versus occupation
-- can be defined as~\cite{dabo:NK}
\begin{eqnarray}
\Pi_{\sigma i}(f)&=&
\Delta E^{\rm Koopmans}_{\sigma i}(f)-\Delta E_{\sigma i}\nonumber\\
&=&\int_0^{f_{\sigma i}}
\left[\epsilon_{\sigma i}(f)-\epsilon_{\sigma i}(f')\right]\,df'\,.
\label{eq:Pi}
\end{eqnarray}
From this result, the SIC energy term based on the generalized
Koopmans theorem has been defined as
\begin{equation}
E_{\rm SIC}^{\rm NK}[\{\rho_{\sigma i}\}]=
\sum_\sigma \sum_{i=1}^N\Pi_{\sigma i}(f_{\rm ref})\,,
\label{eq:NKSIC}
\end{equation}
where $f_{\rm ref}$ is a reference occupation factor
(for many-electron systems,
$f_{\rm ref}=\frac{1}{2}$ was shown to be the best choice~\cite{dabo:NK}).

The total energy versus (fractional) number of electrons relation
calculated by exact DFT should be piecewise linear
with slope discontinuities at integral electron occupations~\cite{piecewise};
however, within the LDA, this energy versus occupation relation is
piecewise convex~\cite{cohen_insights_2008}.
The LDA deviation from the piecewise linearity
is the main reason for the failures of approximate DFTs~\cite{cohen_insights_2008}.
The new SIC functional [Eq.~(\ref{eq:NKSIC})] is introduced to cure this
pathology and to recover the piecewise linearity of exact DFT~\cite{dabo:NK}.
The (bare) NK SIC discussed above and its screened version
explain some of the most important material properties
such as ionization energy and
electron affinity better than PZ SIC.
We refer the reader to Ref.~\cite{dabo:NK}
for the details of NK SIC.

\subsection{B. Implementation}
In order to implement a variational minimization of the total energy
functional,
we adopt the same strategy as the ensemble-DFT approach~\cite{eDFT},
decoupling the dynamics of orbital rotations in the occupied subspace
and that of orbital evolution in the manifold orthogonal to the occupied
subspace. In explicit terms, we minimize the SIC energy through
\begin{equation}
\min_{\{\psi'_{\sigma i}\}}E_{\rm SIC}[\{\psi'_{\sigma i}\}]
=\min_{\{\psi_{\sigma i}\}}\left(\min_{\{O_{\sigma}\}}E_{\rm SIC}[\{\psi_{\sigma i}\},\{O_\sigma\}]\right)\,,
\label{eq:twoloop}
\end{equation}
where $\{\psi'_{\sigma i}\}$ and $\{\psi_{\sigma i}\}$ are
connected by an orthogonal transformation $\{O_{\sigma}\}$ [Eq.~(\ref{eq:U})].
Minimization over the basis orbital wavefunctions $\{\psi_{\sigma i}\}$
and that over the orthogonal transformation
$\{O_{\sigma i}\}$ -- inside the round parenthesis in
Eq.~(\ref{eq:twoloop}) -- correspond to
the outer-loop minimization and inner-loop minimization,
respectively, i.\,e.\,,
given the orbital wavefunctions,
an optimal orthogonal transformation is searched and then
the orbital wavefunctions are evolved.  This process is
repeated until convergence.
Ensemble-DFT minimization has also been discussed
in studying the SIC problem by Stengel and Spaldin~\cite{stengel_spaldin}
and by Kl\"upfel, Kl\"upfel, and J\'onsson~\cite{klupfel}.

The main focus here is on inner-loop minimization.
The gradient matrix $G_{\sigma ij}=\partial E_{\rm SIC}/\partial A_{\sigma ij}$
in Eq.~(\ref{eq:pederson}) is
antisymmetric and real; hence, $-i\, G_{\sigma }$ is Hermitian
(and purely imaginary). Therefore, $-i\,G_\sigma$ can be
diagonalized as
\begin{equation}
-i\,G_{\sigma}=U_\sigma^\dagger\, D_\sigma\, U_\sigma\,,
\end{equation}
or,
\begin{equation}
G_{\sigma}=i\,U_\sigma^\dagger\, D_\sigma\, U_\sigma\,,
\label{eq:G}
\end{equation}
where $U_\sigma$ is a unitary matrix and
\begin{equation}
D_{\sigma ij}=\lambda_{\sigma i}\,\delta_{ij}
\label{eq:D}
\end{equation}
a real diagonal matrix.
From Eq.~(\ref{eq:G}),
we evolve the matrix $A_\sigma$ along the energy gradient with a
step of size $l$
\begin{equation}
\Delta A_{\sigma}=-l\,G_\sigma
=-i\,l\,U_\sigma^\dagger\,D_\sigma\,U_\sigma\,,
\label{eq:dA}
\end{equation}
calculate the updated orthogonal matrix
\begin{equation}
O_\sigma=e^{\Delta A_{\sigma}}=
U_\sigma^\dagger\,e^{-i\,l\,D_\sigma}\,U_\sigma\,,
\label{eq:O_update}
\end{equation}
and then transform the wavefunctions accordingly.

Here, we use the steepest-descent method for the inner-loop minimization.
But one could employ other methods such as damped dynamics
or conjugate gradients.
In each of the inner-loop steps, we evaluate the SIC energy
with two different sets of wavefunctions: first
by using the given wavefunctions [$E_{\rm SIC}(l=0)$]
and second by using the wavefunctions transformed by $O_\sigma$
in Eq.~(\ref{eq:O_update}) with a trial step $l=l_{\rm trial}$
[$E_{\rm SIC}(l=l_{\rm trial})$].
In addition, the gradient at $l=0$ reads
\begin{eqnarray}
\left.\frac{dE_{\rm SIC}(l)}{dl}\right|_{l=0}
&=&\frac{1}{2}\sum_{\sigma i j}\left[\frac{\partial E_{\rm SIC}}{\partial A_{\sigma ij}}\,
\frac{d\Delta A_{\sigma i j}}{dl}\right]_{l=0}\nonumber\\
&=&-\frac{1}{2}\sum_{\sigma ij}|G_{\sigma ij}|^2
\,,
\label{eq:dEdl}
\end{eqnarray}
where we have used Eqs.~(\ref{eq:pederson}) and~(\ref{eq:dA}),
and the fact that only half of the matrix elements of $G_\sigma$
are independent.
Thus, knowing
$E_{\rm SIC}(l=0)$, $E_{\rm SIC}(l=l_{\rm trial})$, and $dE_{\rm SIC}(l)/dl|_{l=0}$,
we can fit a parabola to $E_{\rm SIC}(l)$,
yielding the optimal step $l=l_{\rm optimal}$
and the energy minimum $E_{\rm SIC}(l=l_{\rm optimal})$.
This completes one inner-loop iteration.  We then
use the transformed wavefunctions
to calculate the gradient [Eq.~(\ref{eq:pederson})]
and repeat iterations until the SIC energy converges.

For optimal convergence, we set the step size
based on the highest frequency component
of the gradient matrix, i.\,e.\,,
\begin{equation}
l=\gamma\,l_{\rm c}\,\,\,\,\,\,\left(l_{\rm c}=\frac{\pi}{\lambda_{\max}}\right)\,,
\label{eq:l_c}
\end{equation}
where $\gamma$ is a constant of order $\sim 0.1$ and
$\lambda_{\rm max}$ the maximum eigenvalue of $D_\sigma$, 
\begin{equation}
\lambda_{\max}=\,\max_{\sigma i}\, \lambda_{\sigma i}\,.
\label{eq:lambda_max}
\end{equation}
The critical step $l_{\rm c}$ should be considered
as the point when the transformed wavefunctions
become appreciably different from the original wavefunctions.
Therefore, when we evolve wavefunctions by using a step much
larger than $l_{\rm c}$, a fitting of $E_{\rm SIC}$ versus
$l$ by a parabola will not be successful.
Imposing the constraint $l=\gamma\,l_{\rm c}$ [Eq.~(\ref{eq:l_c})]
when necessary is the key part of our method:
(i) We set the trial step of the first iteration of the inner-loop minimization
according to Eq.~(\ref{eq:l_c}).  (In subsequent iterations, the trial step
$l_{\rm trial}$ is set based on the optimal step of the previous iteration:
we set it to be twice the optimal step
of the previous iteration.)
By setting the initial trial step based on the eigenspectrum
of the gradient matrix, we make
the inner-loop process unaffected by the absolute
magnitude of the SIC energy gradient with respect to
the orthogonal transformation [Eq.~(\ref{eq:pederson})].
(ii) When the calculated optimal
step is larger than $\gamma\,l_{\rm c}$, we set $l_{\rm optimal}=\gamma\,l_{\rm c}$.
This procedure has proven to be instrumental when $E_{\rm SIC}(l)$ versus $l$
relation cannot be fitted well by a parabola.
In such cases, the calculated $l_{\rm optimal}$ can be much larger
than $l_{\rm c}$.
A similar scaling method based on the highest frequency component
of the gradient matrix was used in finding the
MLWFs~\cite{marzari_vanderbilt,Mostofi2008685}.

\subsection{C. MLWFs as an initial guess for the wavefunctions}

SIC tends to localize the orbital wavefunctions [note e.\,g.\,, that
the Hartree
term in Eq.~(\ref{eq:E_SIC_PZ}) will be more negative if the state becomes
more localized].  Therefore, it is natural to consider using some localized
basis functions as an initial guess for the wavefunctions of density
functionals with SIC.
To this end, employing
MLWFs~\cite{marzari_vanderbilt,souza_marzari_vanderbilt}
represents a very promising initial-guess strategy.
Although the possibility of using MLWFs in this regard was
recently suggested~\cite{tsemekhman},
no literature is available on the merit of that scheme.
We address this issue in conjunction with the inner-loop minimization
method discussed in the previous subsection.

\subsection{D. Computational details}

We performed DFT calculations with norm-conserving
pseudopotentials~\cite{TroullierMartins91}
in the LDA~\cite{perdew_zunger} using
the Car-Parrinello (CP) code of the Quantum ESPRESSO
distribution~\cite{baroni:2006_Espresso}
with the inner-loop minimization described in
the previous subsections, and a conventional damped dynamics algorithm
for the outer-loop minimization.
We have performed calculations on both PZ SIC~\cite{perdew_zunger}
and NK SIC~\cite{dabo:NK}.
Except for the case of investigating the effect of using
MLWFs as an initial guess for the wavefunctions, we have
used LDA wavefunctions with some arbitrary phases --
they are not LDA eigenstates -- when we start the calculations.

We performed calculations on a rather big molecule,
C$_{20}$ fullerene.
A supercell geometry was used with the minimum distance
between the carbon atoms in neighboring supercells larger
than 6.7~\AA.
The Coulomb interaction is truncated to prevent
spurious interaction between periodic replicas in different
supercells~\cite{dabo:Coulomb,dabo:arXiv}.

\section{III. Results and Discussion}

\begin{figure*}
  \includegraphics[width=1.6\columnwidth]{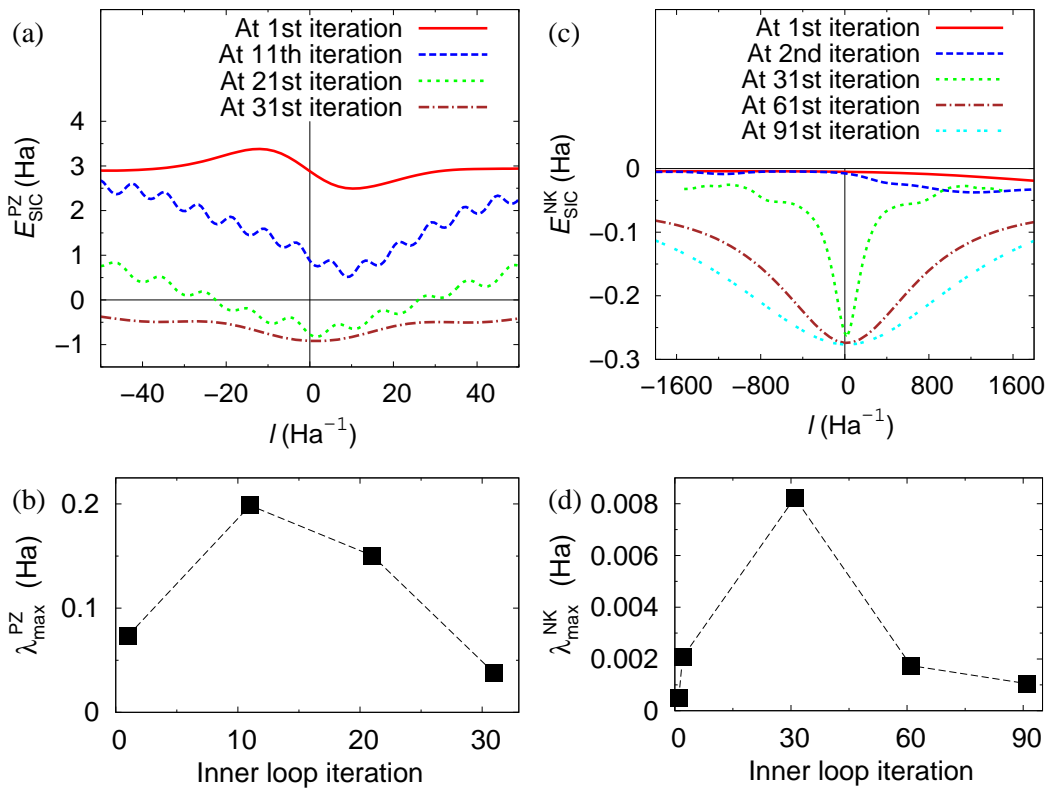}
\caption{(a) Unitary variant part of the total energy within PZ SIC,
$E_{\rm SIC}^{\rm PZ}$ [Eq.~(\ref{eq:E_total})],
for C$_{20}$ versus
step size $l$ [see Eq.~(\ref{eq:dA})] to which
the amount of rotation of the occupied electronic
states is proportional.
(b) Maximum eigenvalue $\lambda^{\rm PZ}_{\rm max}$
[Eq.~(\ref{eq:lambda_max})] of the gradient matrix
$\partial E^{\rm PZ}_{\rm SIC}/\partial A_{ij}$ [Eq.~(\ref{eq:pederson})]
for PZ SIC as a function of the inner-loop iteration steps.
The dashed line is a guide to the eye.
(c) and (d) Similar quantities as in (a) and (b), respectively,
for NK SIC.}
\label{Fig1}
\end{figure*}

In order to find an optimal strategy for the minimization
of SIC DFT, it is important to know how the energy varies
with orthogonal transformations.
We first show the energy variation along the direction
in the orthogonal transformation space parallel to the gradient
[Eq.~(\ref{eq:pederson})] of the energy, i.\,e.\,,
$E_{\rm SIC}(l)$ versus $l$,
where $l$ is a step representing the amount of
orthogonal rotation as defined in Eq.~(\ref{eq:dA}).
Figure~\ref{Fig1}(a) shows the results for PZ SIC
at a few different stages during the inner-loop minimization.
What we can see is that initially $E_{\rm SIC}^{\rm PZ}(l)$
varies slowly with $l$, and then, in the middle
of the inner-loop minimization, varies
fast and then, toward the end of the minimization,
varies slowly again.  There is no good length scale
of $l$ which can consistently describe the variation
of $E_{\rm SIC}(l)$ during the entire process of an inner-loop
minimization.  The speed of the energy variation
at different stages of the inner-loop minimization with
respect to $l$ near $l=0$ can however be very well explained
by $\lambda_{\rm max}$ [Eq.~(\ref{eq:lambda_max})],
which is the fastest frequency component of the gradient matrix
[Eq.~(\ref{eq:pederson})], as shown in Fig.~\ref{Fig1}(b).

We can draw similar conclusions for NK SIC as
shown in Figs.~\ref{Fig1}(c) and~\ref{Fig1}(d).
However, there are a few points that are worth mentioning.
First, the magnitude of NK SIC energy is several times
smaller than that of PZ SIC energy [Figs.~\ref{Fig1}(a) and~\ref{Fig1}(c)].
Second, $\lambda_{\rm max}$, or the main driving force for
orthogonal transformation near $l=0$, for NK SIC is also
much smaller than that for PZ SIC, although eventually both
of them converge to zero at energy minima.
Because of these differences between different SIC functionals,
it is clear that determining the trial step $l_{\rm trial}$
based on $\lambda_{\rm max}$ will be very useful, even more so because
$\lambda_{\rm max}$ is also affected by the arbitrary initial phases
of the wavefunctions,
as will be discussed later (Fig.~\ref{Fig6}).

\begin{figure}
  \includegraphics[width=0.8\columnwidth]{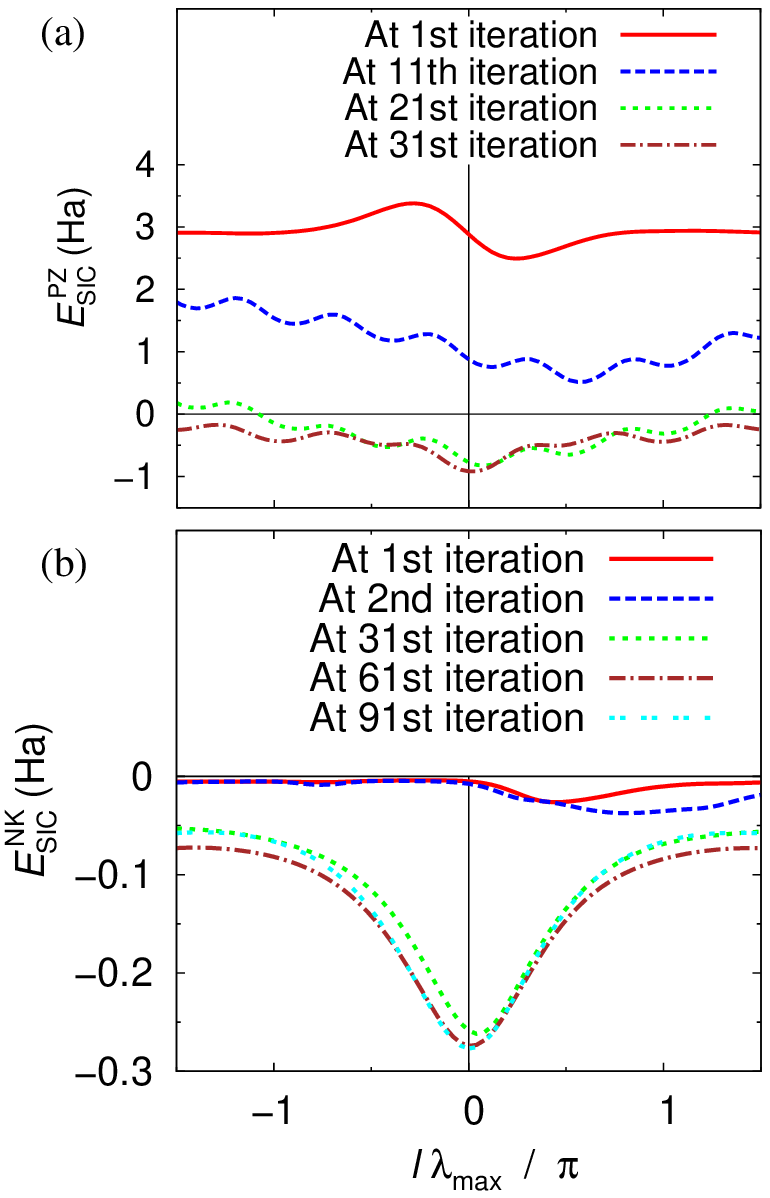}
\caption{
(a) Unitary variant part of the total energy within PZ SIC,
$E_{\rm SIC}^{\rm PZ}$ [Eq.~(\ref{eq:E_total})],
versus $l\,\lambda_{\rm max}/\pi$
[Eq.~(\ref{eq:lp})] for C$_{20}$ at a few
inner-loop iteration steps.
(b) Similar quantity as in (a) for NK SIC.}
\label{Fig2}
\end{figure}

Based on the previous discussion, we now show,
in Fig.~\ref{Fig2}, $E_{\rm SIC}(l)$
as a function of the scaled step
\begin{equation}
l_{\rm scaled}\,\equiv\,{l}\,/\,{l_{\rm c}}\,,
\label{eq:lp}
\end{equation}
i.\,e.\,,
$l$ in units of $l_{\rm c}$.
For both PZ SIC and NK SIC, the energy variation
length scale
near $l=0$ through the entire process of the inner-loop
minimization is $\sim0.5$ in units of $l_{\rm scaled}$.
The results confirm
that indeed a natural length scale for $l$ that should be
used in the inner-loop minimization is the $l_{\rm c}$
defined in Eq.~(\ref{eq:l_c}).
One more thing to note here is that in both PZ SIC and
NK SIC, at the initial stages of the inner-loop iterations,
the energy profile cannot be well fitted by a parabola.
This trend is dramatic especially for NK SIC,
where the $E_{\rm SIC}(l)$ versus $l$ (or $l_{\rm scaled}$)
relation is concave, not convex, at $l=0$.

\begin{figure}
  \includegraphics[width=0.8\columnwidth]{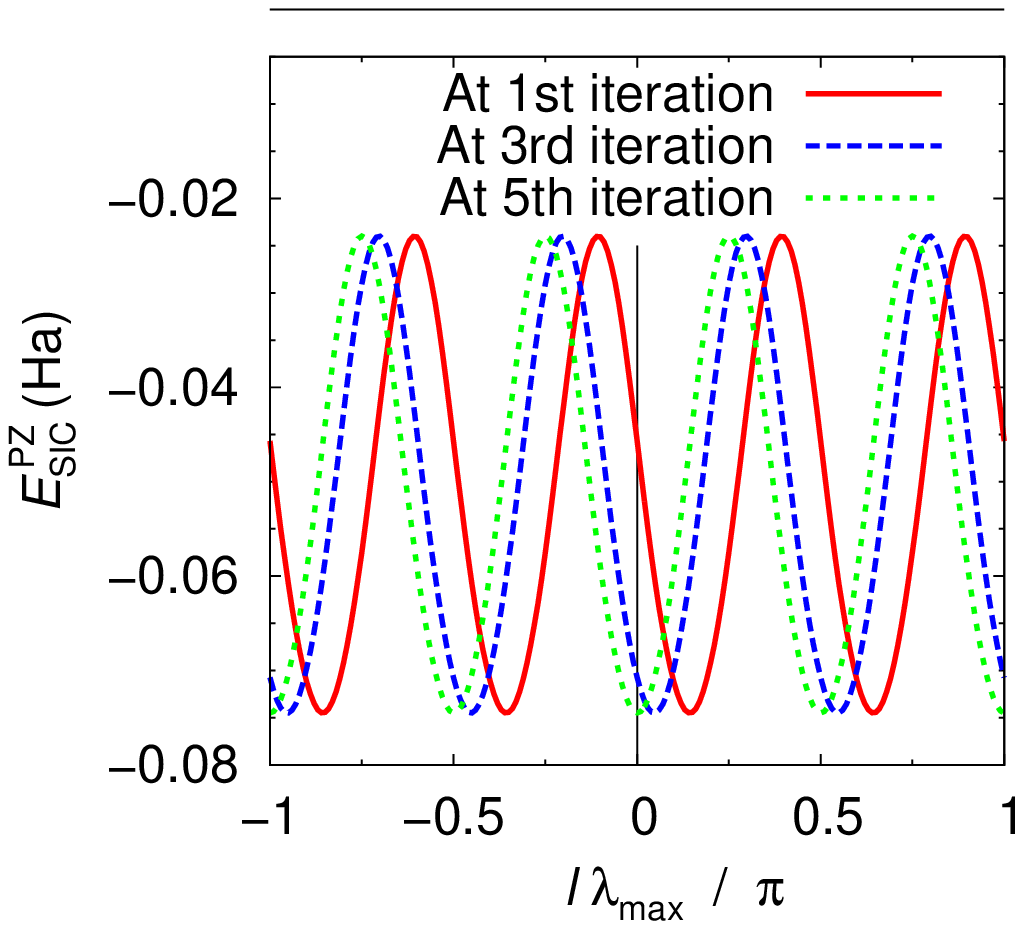}
\caption{
(a) Unitary variant part of the total energy within PZ SIC,
$E_{\rm SIC}^{\rm PZ}$ [Eq.~(\ref{eq:E_total})], versus
$l\,\lambda_{\rm max}/\pi$ [Eq.~(\ref{eq:lp})]
for a carbon atom at a few inner-loop iteration steps.
}
\label{Fig3}
\end{figure}

This can be best understood using a simple system:
a carbon atom which has, in our pseudopotential
calculations, two orbitals
($2s$ and $2p$), i.\,e.\,, it is a two-level system.
The PZ SIC energy $E_{\rm SIC}^{\rm PZ}(l)$ versus
$l_{\rm scaled}$ is shown in Fig.~\ref{Fig3}.
The profile is sinusoidal with a period of 0.5,
rather than parabolic for the entire process of minimization.
Notably, the period 0.5 in units of $l_{\rm scaled}$
is similar to the previously discussed length scale
for C$_{20}$ fullerene.
The shape of the curve does not change as we proceed
in the inner-loop minimization;
the only variation is that the minimum of
the curve moves toward the origin ($l_{\rm scaled}=0$).

We can understand this behavior as follows. The gradient
matrix in Eq.~(\ref{eq:pederson}) for a carbon atom
is of the form
\begin{equation}
G=\left(
\begin{array}{cc}
0 & c\\
-c & 0
\end{array}\right)=i\,c\,\sigma_y
\,,
\label{eq:grad_C}
\end{equation}
where $c$ is a real constant
and $\sigma_y$ the Pauli matrix.
(We dropped the spin index for obvious reasons.)
Assuming (without losing generality) that $c>0$,
the maximum eigenvalue of $G$ is
\begin{equation}
\lambda_{\rm max}=c
\end{equation}
and the orthogonal transformation matrix
[Eqs.~(\ref{eq:dA}) and~(\ref{eq:O_update})]
is given by
\begin{equation}
O=e^{-lG}=\cos\,(lc)\,I\,-\,i\,\sin\,(lc)\,\sigma_y\,,
\end{equation}
or, using $l_{\rm scaled}$ [Eq.~(\ref{eq:lp})],
\begin{equation}
O=\cos\,(\pi\,l_{\rm scaled})\,I\,-\,i\,\sin\,(\pi\,l_{\rm scaled})\,\sigma_y\,.
\end{equation}
In particular, when $l_{\rm scaled}=0.5$, $O=-i\,\sigma_y$,
and, under this orthogonal transformation $O$, $\psi_1'=-\psi_2$ and $\psi_2'=\psi_1$,
i.\,e.\,, $O$ just exchanges the two orbital wavefunctions
(plus a trivial sign change).
When the original wavefunctions $\psi_1$ and $\psi_2$
correspond to the maximum SIC energy configuration,
the new set of wavefunctions $\psi_1'$ and $\psi_2'$
will correspond also to the SIC energy maximum.
Therefore, the period of $E_{\rm SIC}(l)$ versus $l_{\rm scaled}$
will be 0.5 in agreement with our calculation [Fig.~\ref{Fig3}].
(The shape of the curve is not exactly sinusoidal
and varies slightly with the kind of SIC used.)

For this example,
which part of the sinusoidal-like curve
one starts the inner-loop minimization from
depends on the initial orbital wavefunctions
(and an arbitrary rotation of them).
If we start
from the LDA eigenstates, the SIC energy is at its maximum
(roughly speaking, the LDA eigenstates are the most delocalized and the
SIC energy is highest) and the inner-loop minimization starts from
the top of the sinusoidal-like curve, and hence
(i) the driving force for the orthogonal transformation is extremely
weak (zero at the maximum) and (ii) $E_{\rm SIC}(l)$ versus
$l_{\rm scaled}$
is concave. For these reasons, if we do not properly scale
$l$, or if we do not constrain $l$ during the inner-loop minimization
process, the minimization process based on the assumption that the
energy profile is convex parabolic may become unstable or extremely slow.
This discussion is also relevant to other systems,
as we have seen in the case of C$_{20}$ fullerene.

\begin{figure*}
  \includegraphics[width=1.6\columnwidth]{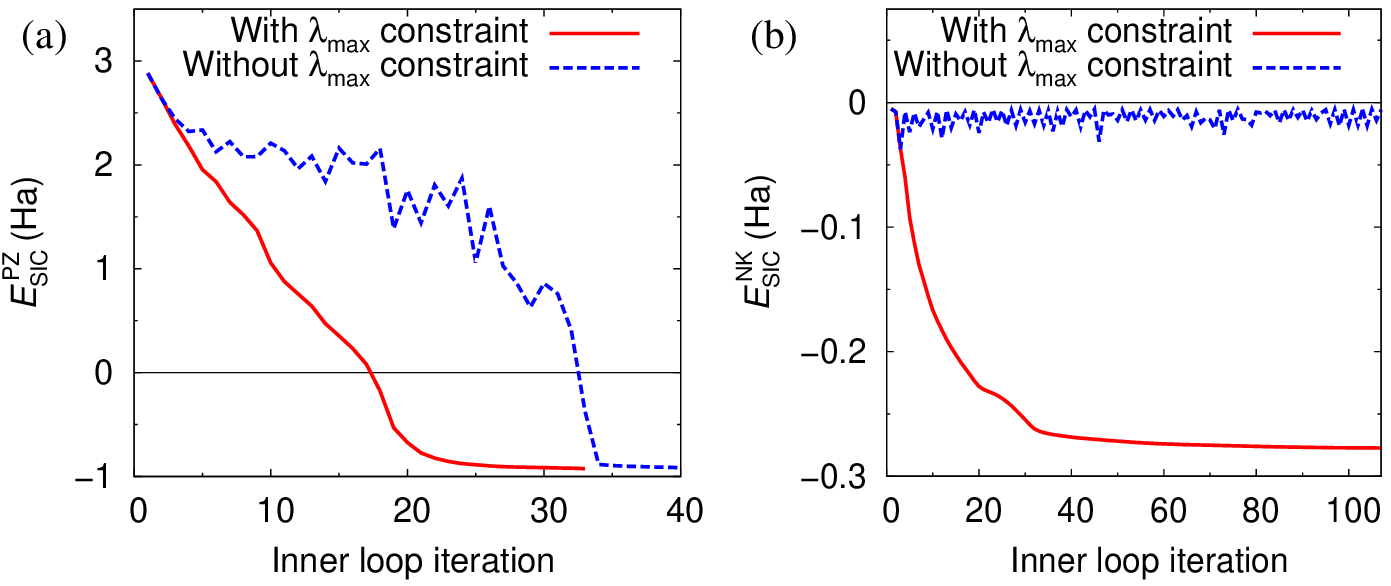}
\caption{
(a) Unitary variant part of the total energy within PZ SIC,
$E_{\rm SIC}^{\rm PZ}$ [Eq.~(\ref{eq:E_total})], versus
the inner-loop iteration step for $C_{20}$ with and
without using the $\lambda_{\rm max}$ constraint
(see text)
during the inner-loop minimization process.
(b) Similar quantity as in (a) for NK SIC.
}
\label{Fig4}
\end{figure*}

Figure~\ref{Fig4}(a) compares the performance of the inner-loop
minimization for the case PZ SIC.  In one case (dashed or blue curve),
we take the
optimal step size $l_{\rm optimal}$ obtained from fitting
$E_{\rm SIC}^{\rm PZ}(l)$ versus $l$ by a parabola
from three calculated quantities: $E_{\rm SIC}^{\rm PZ}(l=0)$,
$E_{\rm SIC}^{\rm PZ}(l=l_{\rm trial})$, and
$dE_{\rm SIC}^{\rm PZ}(l)/dl|_{l=0}$.
In the other case (solid or red curve),
if the calculated $l_{\rm optimal}$ is larger than $\gamma\,l_{\rm c}$
(with $\gamma=0.1$) [Eq.~(\ref{eq:l_c})],
we set $l_{\rm optimal}=\gamma\,l_{\rm c}$.
Apparently, by using this constraint based on $l_{\rm c}$,
or, $\lambda_{\rm max}$, the inner-loop minimization process
becomes more stable and faster.
(In both cases, the trial step of the first iteration was set
to $l_{\rm trial}=\gamma\,l_{\rm c}$.)
The difference between using and not using
this $\lambda_{\rm max}$ constraint is dramatic for NK SIC
[Fig.~\ref{Fig4}(b)].
This again is due to (i) the small gradient of the SIC energy
with respect to the variation of the orthogonal transformation,
and (ii) non-concave-parabolic dependence of
$E_{\rm SIC}(l)$ on $l$.

\begin{figure*}
  \includegraphics[width=1.6\columnwidth]{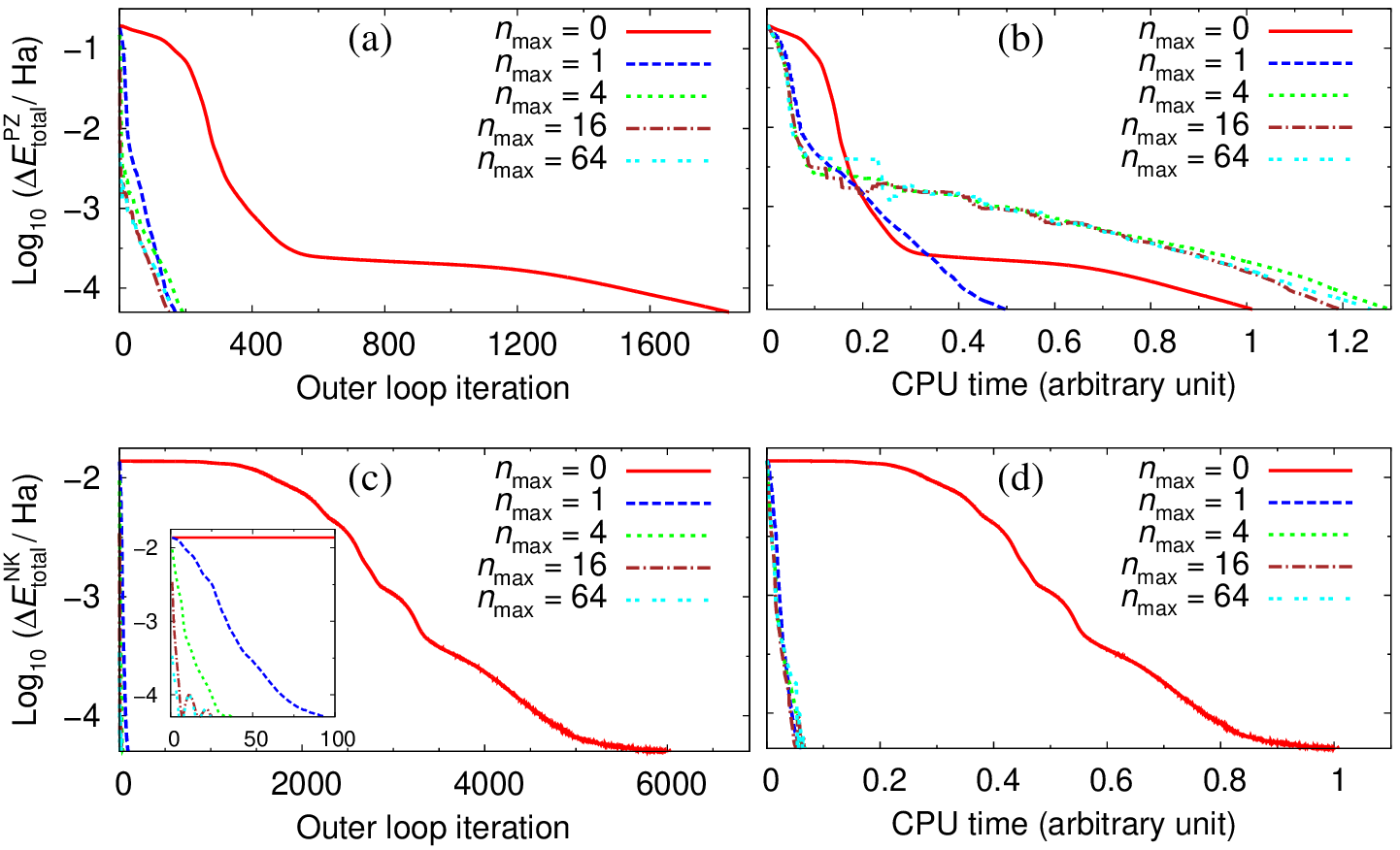}
\caption{
(a) Logarithm of the difference between the
total energy per carbon atom at each outer-loop iteration step and
that at convergence for PZ SIC,
$\log\,\Delta E_{\rm total}^{\rm PZ}$, at a few different values
of $n_{\rm max}$.  Here, $n_{\rm max}$ is the maximum number of inner-loop
iteration steps performed in one outer-loop iteration step,
$n_{\rm max}=0$ being the case without inner-loop minimization.
(b) ${\rm Log}\,\Delta E_{\rm total}^{\rm PZ}$ versus the CPU time.
(c) and (d) Similar quantities as in (a) and (b), respectively,
for NK SIC.
}
\label{Fig5}
\end{figure*}

Until now, our focus was on the inner-loop minimization. Now
we look at the entire minimization procedure including the outer loop.
In order to find an optimal minimization strategy, we have performed
our calculations by restricting the number of inner-loop minimization
iterations per each outer-loop iteration to be less than or
equal to $n_{\rm max}$.
(However, not every outer-loop iteration will require $n_{\rm max}$
inner-loop iterations because the SIC energy may be converged
earlier during inner-loop minimization. We exit the inner loop
if the energy difference between consecutive iterations is lower
than the energy convergence threshold of $10^{-5}$~Ry.)
The case without inner-loop minimization is denoted by $n_{\rm max}=0$.
Figure~\ref{Fig5}(a) shows the convergence of PZ SIC energy
for various different
choices of $n_{\rm max}$. In all cases where the inner-loop minimization
routine is used (i.\,e.\,, $n_{\rm  max}>0$),
the total number of outer-loop iterations necessary
to achieve the same level of convergence is much smaller than
that when no inner-loop minimization is used.
This, however, does not necessarily mean that the total computation
time is reduced.  In Fig.~\ref{Fig5}(b), we show the CPU
time dependence of the SIC energy (the results include both the
inner-loop and outer-loop minimization iterations).
Surprisingly, in all cases other than $n_{\rm max}=1$,
inner-loop minimization actually slows down the computation
for PZ SIC. When we set $n_{\rm max}=1$, i.e., if
the number of inner-loop iterations per each outer-loop iteration
is restricted to~1, we find about twice improvement over when
no inner-loop minimization is performed in terms of
the CPU time.

The case of NK SIC is very different.
Figures~\ref{Fig5}(c) and~\ref{Fig5}(d) shows that
inner-loop minimization reduces not only the required
number of outer-loop iterations but also the CPU time significantly.
Especially, the CPU time is reduced by $\sim20$ times when we perform
inner-loop minimization, and is rather insensitive to $n_{\rm max}$.

These results on PZ SIC and NK SIC support that the presented
method works regardless of the absolute magnitude of
the SIC energy gradient with respect to the orthogonal transformation
[Eq.~(\ref{eq:pederson})].
The method can be applied to density functionals with other kinds
of SIC.  For example, SIC with
screening, for which the
total energy is given by
\begin{equation}
E_{\rm total}=E_{\rm LDA}+\alpha\,E_{\rm SIC}\,\,\,(\alpha<1)\,,
\end{equation}
will have the SIC energy gradient lower in magnitude than
the unscreened version of SIC ($\alpha=1$), and our method
will be more useful.

It has to be noted that the relative CPU time among different
calculations shown in Fig.~\ref{Fig5} at different stages
of the minimization is affected only by the ratio of the
CPU time for one inner-loop iteration to that for one outer-loop iteration.
Therefore, the relative CPU time is rather insensitive to the
complexity of the system studied, and in that sense is meaningful.
(The absolute CPU time is also affected much by the complexity of
the system, the performance and number of processors, etc.)
In our case, one inner-loop iteration for PZ SIC
takes 3.6 times as long as one outer-loop iteration
and one inner-loop iteration for NK SIC takes
2.0 times as long as one outer-loop iteration.

\begin{figure*}
  \includegraphics[width=1.6\columnwidth]{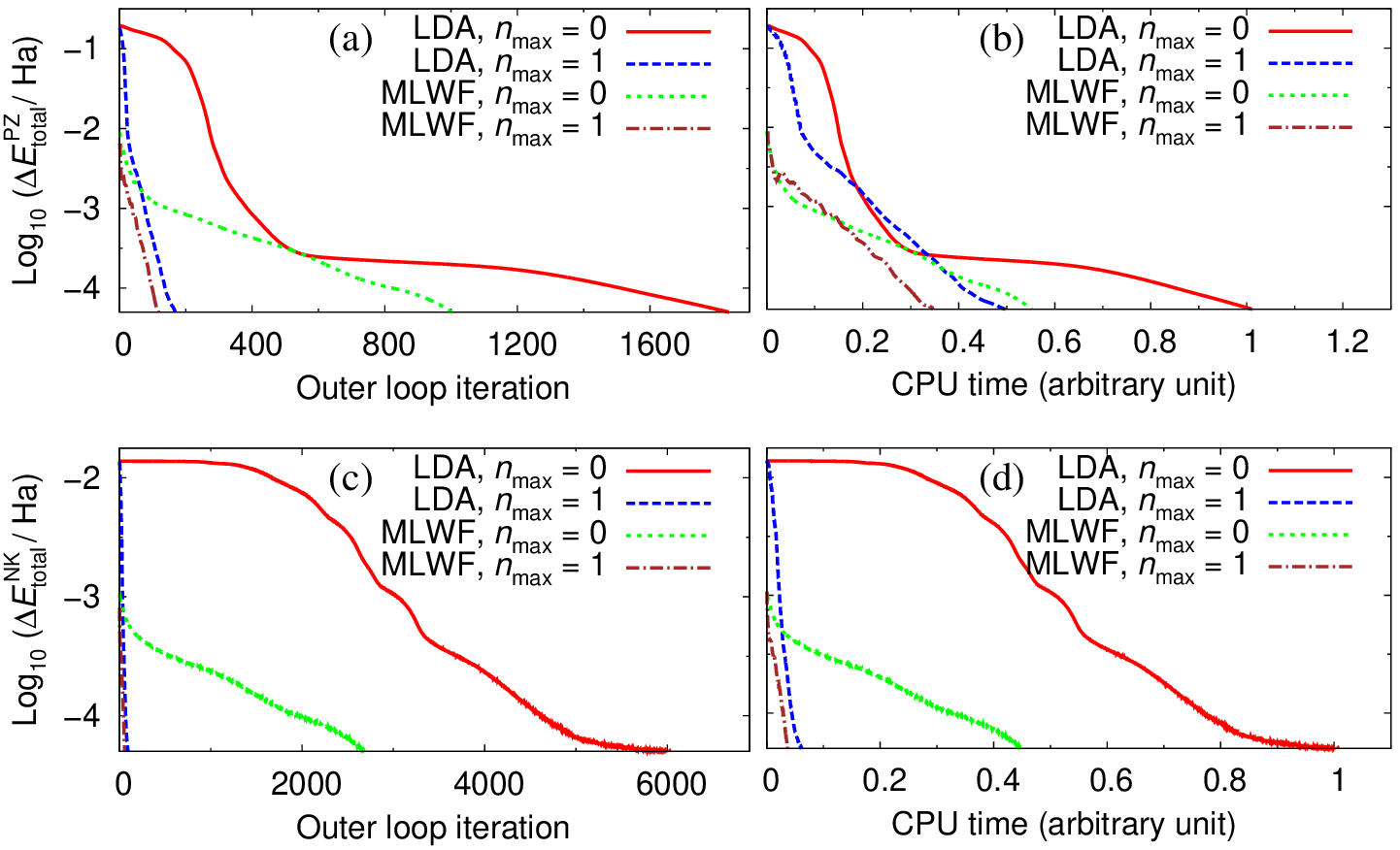}
\caption{
(a) ${\rm Log}\,\Delta E_{\rm total}^{\rm PZ}$ at each outer-loop
iteration step. The curves denoted by LDA show results starting
from the LDA wavefunctions (with arbitrary phases), whereas
those denoted by MLWF show results starting from the maximally
localized Wannier functions obtained from those LDA wavefunctions.
Also, $n_{\rm max}$ is the maximum number of inner-loop
iteration steps performed in one outer-loop iteration step,
$n_{\rm max}=0$ being the case without inner-loop minimization.
(b) ${\rm Log}\,\Delta E_{\rm total}^{\rm PZ}$ versus the CPU time.
(c) and (d) Similar quantities as in (a) and (b), respectively,
for NK SIC.
}
\label{Fig6}
\end{figure*}

Finally, we discuss how useful it is to use
MLWFs~\cite{marzari_vanderbilt,souza_marzari_vanderbilt}
as an initial guess for
the wavefunctions~\cite{klupfel}.
The following description is relevant for both
PZ SIC [Figs.~\ref{Fig6}(a) and~\ref{Fig6}(b)] and
NK SIC [Figs.~\ref{Fig6}(c) and~\ref{Fig6}(d)]
and whether or not the inner-loop minimization is employed.
Figure~\ref{Fig6} shows that when MLWFs are used,
the initial total energy is lower than when LDA wavefunctions with
arbitrary phases is used.  On the other hand, the slope of
$\log[({\rm current\,\,\,total\,\,\, energy})-({\rm converged \,\,\,total\,\,\, energy})]$
versus either the number of outer-loop iterations
or the relative CPU time is not very different in the two cases.
Therefore, it is advantageous to use MLWFs as an initial guess
for the wavefunctions; however, the lower the energy convergence
threshold the smaller the relative advantage.

\section{IV. Conclusions}

In summary, we have developed a variational, stable and
efficient approach for the total-energy
minimization of unitary variant functionals,
as they appear in self-interaction corrected formulations,
with a focus
on properly minimizing the energy by unitary transformations
of the occupied manifold.
In particular, we have shown that the energy changes along the
gradient direction can be very different from being convex parabolic,
and suggested the use of the maximum frequency component of
the gradient matrix in determining optimal rotations for
the inner-loop minimization.
When maximally localized Wannier functions are used as an initial
guess for the wavefunctions,
the initial energy decreases significantly from that
corresponding to wavefunctions with arbitrary phases; however, the
logarithmic energy convergence rate remains similar in the two cases.
We expect that the results will be useful for investigating
the physical properties of complex materials and big molecules
with self-interaction corrected density functional theory.

We thank fruitful discussions with Peter Kl\"upfel and Simon Kl\"upfel.
CHP acknowledges financial support from
Intel Corporation.

\end{document}